%% file: revtex4.tex
\documentclass[pre,twocolumn,twoside,byrevtex,superscriptaddress,floatfix]{revtex4-1}

\usepackage{settings}

\setboolean{twocolswitch}{true}
\begin{document}

\title{\protect \input{title}}
\input{authors}

\date{\today}

\begin{abstract}
  \protect
  \input{abstract.tex} 
\end{abstract}

\input{pacs}

\maketitle


\section{Introduction}\label{sec:introduction} 
\input{sections/introduction.tex}

\section{Data}\label{sec:data}
\input{sections/data.tex}

\section{Results}\label{sec:results}
\input{sections/results.tex}

\section{Discussion}\label{sec:discussion}
\input{sections/conclusion.tex}

\section{Methods}\label{sec:methods}
\input{sections/methods.tex}

\acknowledgments 
\input{sections/acknowledgements.tex} 

\bibliography{references}

\clearpage

\newwrite\tempfile
\immediate\openout\tempfile=startsupp.txt
\immediate\write\tempfile{\thepage}
\immediate\closeout\tempfile

\setcounter{page}{1}
\renewcommand{\thepage}{S\arabic{page}}
\renewcommand{\thefigure}{S\arabic{figure}}
\renewcommand{\thetable}{S\arabic{table}}
\setcounter{figure}{0}
\setcounter{table}{0}

\appendix
\input{sections/supplementary.tex}

\end{document}

%% file: title.tex
Divergent modes of online collective attention to the COVID-19 pandemic
are associated with future caseload variance

%% file: authors.tex
\author{
\firstname{David Rushing}
\surname{Dewhurst}
}
\affiliation{
  Vermont Complex Systems Center,
  Computational Story Lab,
  The University of Vermont,
  Burlington, VT 05405
} 
\affiliation{
MassMutual Data Science,
Boston, MA 02110
}

\author{
\firstname{Thayer}
\surname{Alshaabi}
}
\thanks{These authors contributed equally to this work.}
\affiliation{
  Vermont Complex Systems Center,
  Computational Story Lab,
  The University of Vermont,
  Burlington, VT 05405
} 

\author{
\firstname{Michael V.}
\surname{Arnold}
}
\thanks{These authors contributed equally to this work.}
\affiliation{
  Vermont Complex Systems Center,
  Computational Story Lab,
  The University of Vermont,
  Burlington, VT 05405
} 
\author{
\firstname{Joshua R.}
\surname{Minot}
}
\thanks{These authors contributed equally to this work.}
\affiliation{
  Vermont Complex Systems Center,
  Computational Story Lab,
  The University of Vermont,
  Burlington, VT 05405
} 
\author{
\firstname{Christopher M.}
\surname{Danforth}
}
\affiliation{
  Vermont Complex Systems Center,
  Computational Story Lab,
  The University of Vermont,
  Burlington, VT 05405
} 
\affiliation{
Department of Mathematics and Statistics,
The University of Vermont, 
Burlington, VT 05405
}
\author{
\firstname{Peter Sheridan}
\surname{Dodds}
}
\affiliation{
  Vermont Complex Systems Center,
  Computational Story Lab,
  The University of Vermont,
  Burlington, VT 05405
} 
\affiliation{
Department of Mathematics and Statistics,
The University of Vermont, 
Burlington, VT 05405
}


%% file: abstract.tex
Using a random 10\% sample of tweets authored from 2019-09-01 through 2020-04-30, we analyze the dynamic behavior of words (1-grams) used on Twitter to describe the ongoing COVID-19 pandemic.
Across 24 languages, we find two distinct dynamic regimes: One characterizing the rise and subsequent collapse in collective attention to the initial Coronavirus outbreak in late January, and a second that represents March COVID-19-related discourse.
Aggregating countries by dominant language use, we find that
volatility in the first dynamic regime is associated with future volatility in new cases of COVID-19 roughly three weeks (average 22.49 $\pm$ 3.26 days) later.
Our results suggest that surveillance of change in usage of epidemiology-related words on social media may be useful in 
forecasting later change in disease case numbers,
but we emphasize that our current findings are not causal or necessarily predictive.

%% file: pacs.tex
\pacs{89.65.-s,89.75.Da,89.75.Fb,89.75.-k}


%% file: sections/introduction.tex
COVID-19 is a potentially lethal viral respiratory disease that is causing a global pandemic \cite{wu2020characteristics,world2020coronavirus}.
While Coronavirus testing availability is suboptimal
\cite{cheng2020diagnostic}, social media data can be part of an effective strategy for infectious disease surveillance
\cite{bodnar2013validating,charles2015using,santillana2015combining,sculley2010web,wojcik2020survey,lampos2020tracking}.
Previous work has demonstrated that online collective attention to COVID-19 as measured by social media 
activity has fluctuated from the
date of the first public report of the disease (2019-12-31) to near the time of writing 
(2020-04-30) \cite{dong2020interactive,li2020characterizing,alshaabi2020world}.

In this work we analyze time series of word (1-gram) ranks on Twitter computed from a 10\% random sample of all messages.
We find that the temporal dynamics of this discourse separate into two distinct clusters,
one ($C_1$) that contains words contributing to the explosive rise in online discussion of COVID-19 prevention and treatment during March 2020 and another ($C_2$) that contains words contributing to the
 rise and subsequent fall in collective attention to COVID-19 during mid-January -- mid-February 2020.
Variance of percent changes in word time series closest to the centroid of $C_2$ is a consistent
 leading indicator of variance in percent change in new cases of COVID-19. 
We close with a short discussion of the implications and limitations of these findings, and suggestions for future research
\footnote{
All relevant data, code, and figures will eventually be hosted at 
\url{http://compstorylab.org/covid19ngrams/}.
}.

%% file: sections/data.tex
We analyzed time series of word usage on a random 10\% sample of tweets written between 2019-09-01 and 2020-04-30.
For each language under study, we considered only the top 1000 words used in the language as ranked 
during the first three weeks of March 2020 \cite{alshaabi2020world}, and
restrict our analysis to the same 24 languages analyzed in a previous work.
Languages are detected and annotated using a previously-introduced procedure \cite{alshaabi2020growing}.
We obtained data on languages spoken in each country from the Australian Federal Department of 
Social Services and data on number of new COVID-19 cases by country from the European Centers for Disease
Control
\footnote{
Language list by country is available at 
\url{https://www.dss.gov.au/sites/default/files/files/foi\_disclosure\_log/12-12-13/language-list.pdf}
and new case numbers are available at 
\url{https://www.ecdc.europa.eu/en/publications-data/}}.

%% file: sections/results.tex
\subsection{Divergent modes of COVID-19 related language}

We find $k^* = 6$ clusters of normalized log rank word usage timeseries using the algorithm detailed in 
Sec.\ \ref{sec:cluster-number-selection}.
We compute these clusters using the entire dataset, i.e., aggregating all log rank time series in each of the 24
languages under study.
Of these clusters, two are composed primarily of words that do not appear to relate to COVID-19.
The remaining four clusters contain language that relates to COVID-19 both explicitly and implicitly.
We combine these clusters into two aggregate clusters using the methodology defined in Sec.\ \ref{sec:methods}.
\begin{figure*}
\centering
\includegraphics[width=\textwidth]{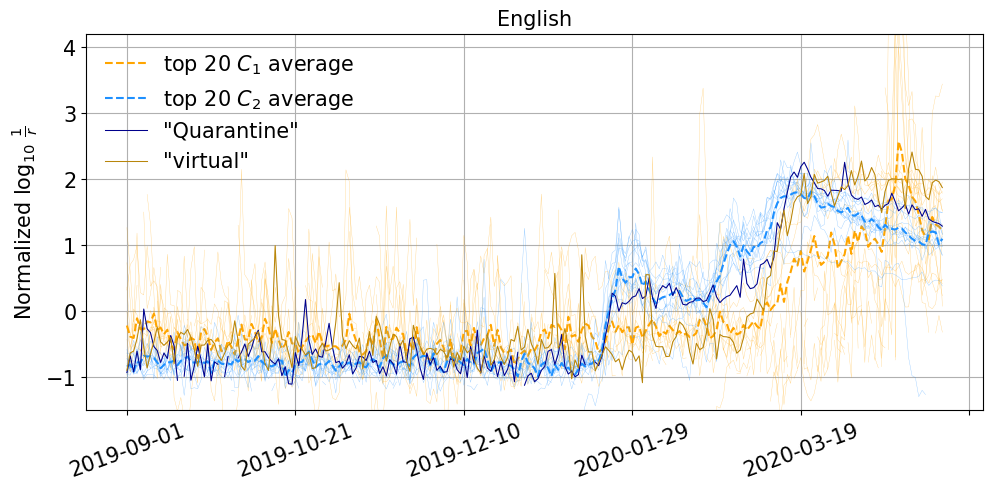}
\caption{
The rise in collective attention to COVID-19 during late January 2020 to early February 2020 followed by a marked 
decline preceding the global pandemic is generated by two distinct clusters of COVID-19 related language.
We display the mean normalized log rank timeseries of the top 20 English words closest
 to each of $E[C_1]$ and $E[C_2]$ in thinner, dashed curves,
and the single English word closest to each of $E[C_1]$ and $E[C_2]$ in thin solid curves.
Increasing granularity (centroids $\rightarrow$ top 20 words $\rightarrow$ single representative word)
is associated with exaggeration of the dynamics of  $E[C_1]$ and $E[C_2]$.
Before normalization we map $\log_{10}r \mapsto \log_{10} \frac{1}{r}$ so that higher values on the vertical axis are lower values of (log) rank.
For comparison, the word ``pandemic'' rises in popularity from a rank of 133,445 on December 21 to a rank of 188 on March 17, while the word ``flatten'' goes from being the 100,913th most popular English word on January 20 to being 2,131st most used word on March 15.
}
\label{fig:cluster-decomp-english}
\end{figure*}
We label these clusters $C_1$ and $C_2$ and their cluster centroids $E[C_1]$ and $E[C_2]$ respectively.
(The ordering of the cluster subscripts comes from the respective maxima of their cluster centroids.)
$E[C_1]$ exhibits very little variation until the first week of March 2020, where it begins a sustained increase
in time.
Conversely, $E[C_2]$ exhibits a smaller increase at the end of January 2020 followed by a larger increase
in the second week of February 2020. This second increase in $E[C_2]$ is followed by another sustained
increase until mid March 2020.
\medskip

\begin{figure*} 
\centering
\includegraphics[width=\textwidth]{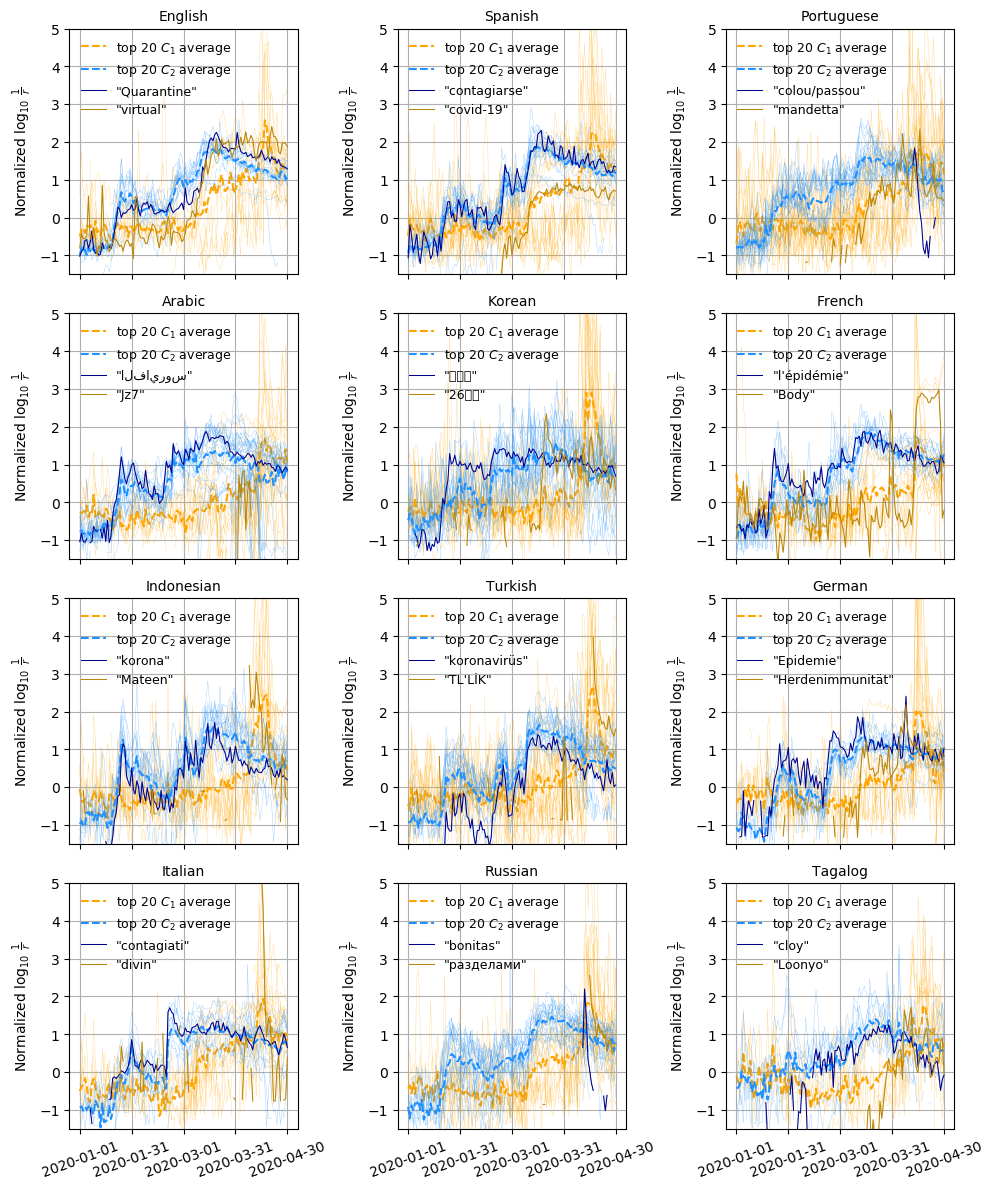}
\caption{
We display the mean normalized log rank timeseries of the top 20 words closest
 to each of $E[C_1]$ and $E[C_2]$ in dashed curves and the single
 word closest to each of $E[C_1]$ and $E[C_2]$ in thin solid curves
 for each of the first 12 of 24 languages.
The divergent modes of dynamic behavior are consistent across most languages, with some languages
(English, French, German, and Indonesian) displaying prominently larger peaks in words closest to $E[C_2]$ during
late January through early February 2020.
Other languages, such as Korean and Tagalog, do not display this behavior.
}
\label{fig:cluster-set0}
\end{figure*}

\begin{figure*}
\centering
\includegraphics[width=\textwidth]{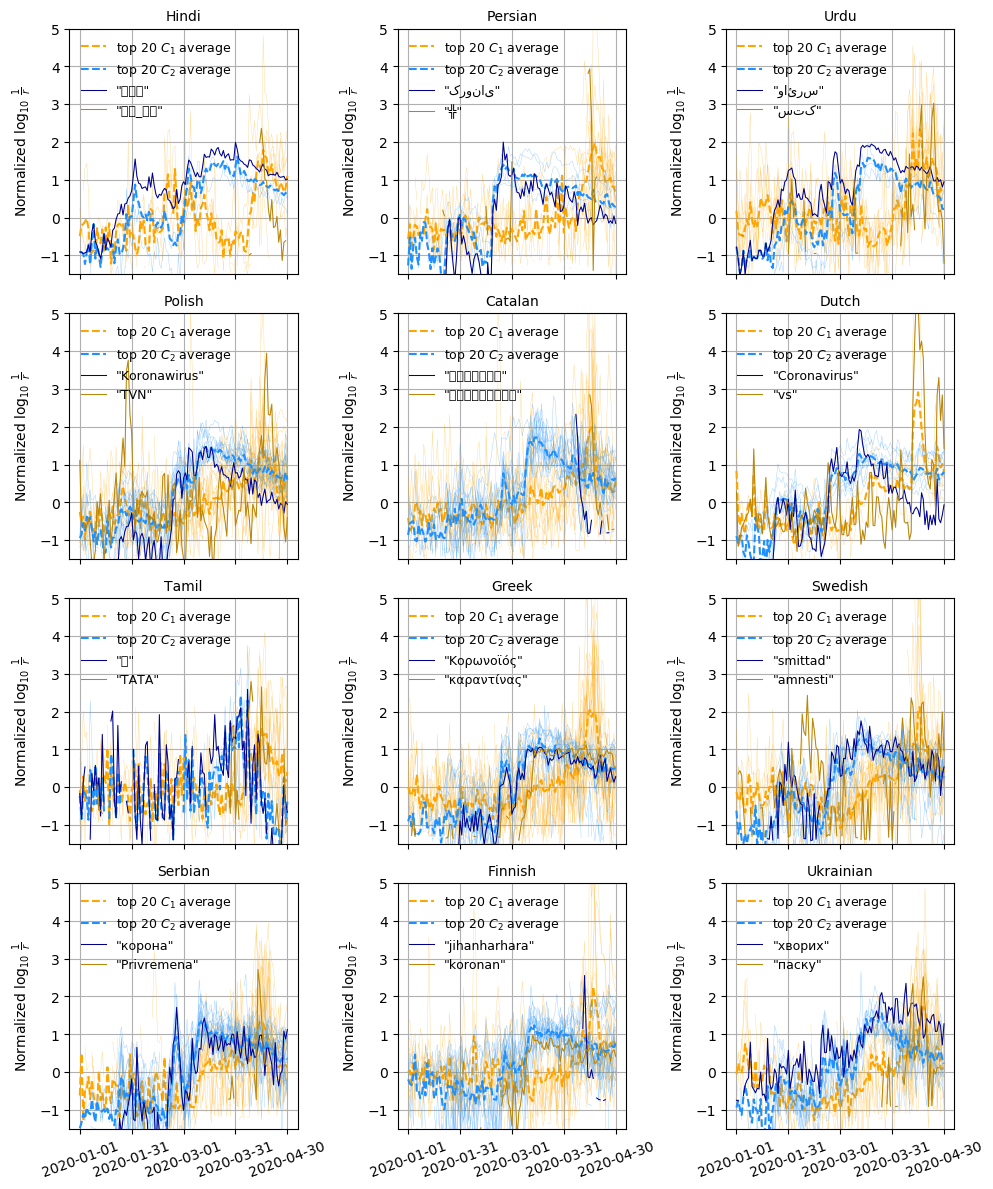}
\caption{
For the second 12 of 24 languages, we display the mean normalized log rank timeseries of the top 20 words closest
 to each of $E[C_1]$ and $E[C_2]$ in dashed curves and the single
 word closest to each of $E[C_1]$ and $E[C_2]$ in thin solid curves.
(We display the first 12 of 24 languages in Fig.\ \ref{fig:cluster-set0}.)
}
\label{fig:cluster-set1}
\end{figure*}

\begin{figure*}
\centering
\includegraphics[width=\textwidth]{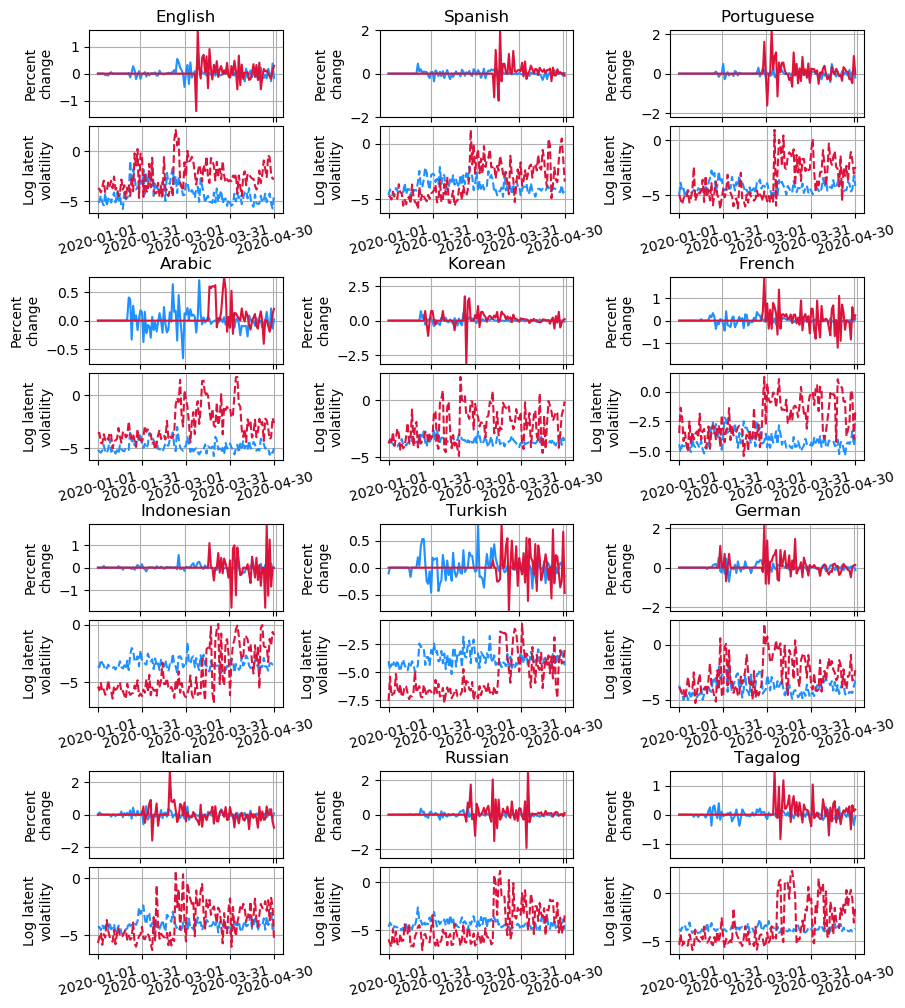}
\caption{
	We display percent-change time series and associated latent log variance (volatility) time series
	for both mean log rank timeseries of the top 20 words closest to $E[C_2]$ (blue curves) and new case number time series (red time series) for each language.
	This figure presents the first 12 of 24 languages.
	There is a positive association between peak volatility in log rank word usage and future peak volatility in new infection case numbers.
	For all languages but four (Swedish, Urdu, Finnish, and Ukrainian), the peak-to-peak
	difference (P2PD) between case and log rank volatility is non-negative.
	The observed P2PD empirical cumulative distribution function (cdf) is not reproduced by a simple difference-of-Poissons (Skellam) model as it exhibits heavier tails than those generated by this model. 
	However, it is reproduced by a Dirichlet process Poisson mixture model. 
	We describe this model in Sec. \ref{sec:methods} in more detail.
}
\label{fig:c5-case0}
\end{figure*}

\begin{figure*}
\centering
\includegraphics[width=\textwidth]{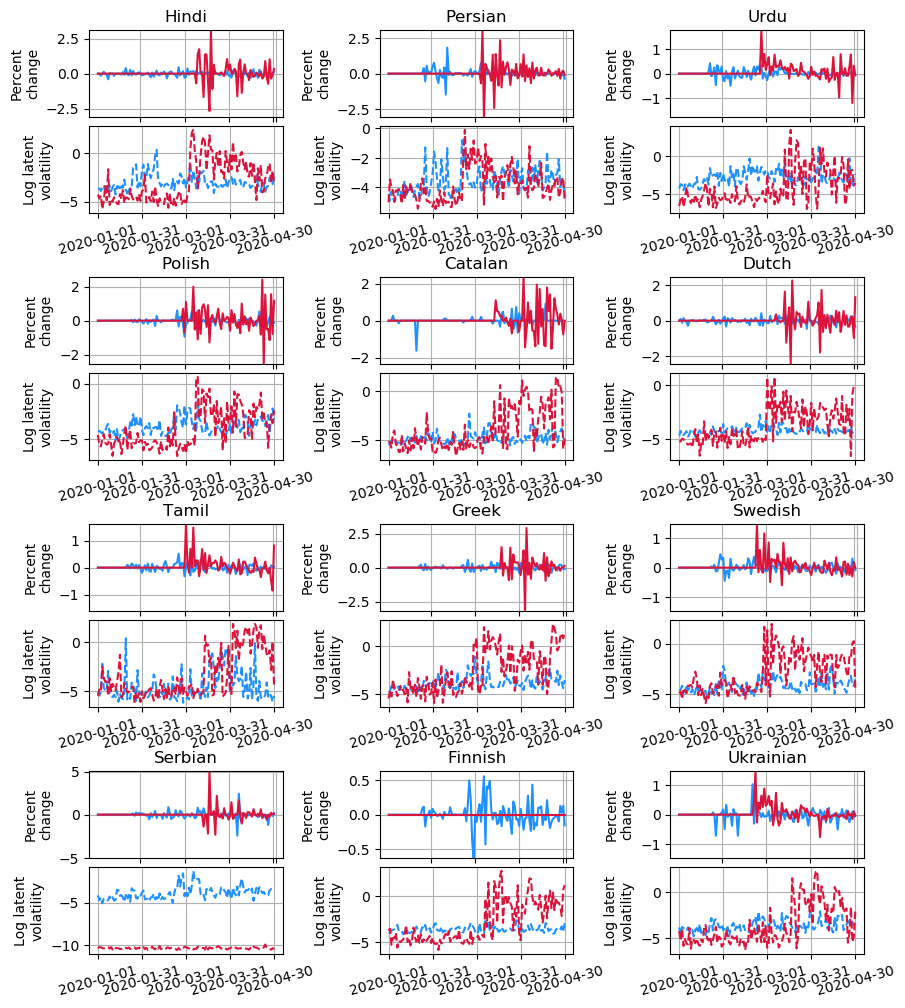}
\caption{
	The second 12 of 24 percent-change and latent volatility time series for log rank (blue time series) and new case load (red time series);
	we display the first 12 of 24 and provide an expanded description in 
	Fig.\ \ref{fig:c5-case0}.
}
\label{fig:c5-case1}
\end{figure*}

This divergent dynamic behavior is amplified when restricting analysis to sets of individual word time series
that are closest to $E[C_1]$ or $E[C_2]$ in Euclidean distance.
The mean normalized log rank timeseries of the top 20 words in each language that were closest to
$E[C_1]$ and $E[C_2]$ exhibit the same qualitative behavior for most of the 24 languages under study, 
but this behavior is amplified (greater magnitudes of increase and decrease).
We display these dynamics for English in Fig.\ \ref{fig:cluster-decomp-english} and for all 24 languages under 
study in Figs.\ \ref{fig:cluster-set0} and \ref{fig:cluster-set1}. 
We plot languages in order of frequency of usage on Twitter in Figs.\ 
\ref{fig:cluster-set0}, \ref{fig:cluster-set1}, \ref{fig:c5-case0},
and \ref{fig:c5-case1}.
For interpretable visualization, we invert ranks ($\log_{10}r \mapsto \log10 \frac{1}{r}$) before normalization and before plotting, so that lower ranked words --- words that are more popular and are receiving more attention --- are higher on the vertical axis than words of higher rank corresponding to lower popularity.
We display the top 20 words associated with $C_1$ and $C_2$ in 
Tab.\ \ref{tab:top20-english}.
\begin{table}
\centering
\begin{tabular}{|c|c|} \hline
   Top 20 closest to $E[C_1]$  & Top 20 closest to $E[C_2]$  \\ \hline
virtual & Quarantine\\
villain's & quarantine\\
Starmer & Virus \\
lockdown & Corona \\
LIBERATE & quarantined \\
\$24,000 & pandemics \\
pallbearers & Fauci \\
Hydroxychloroquine & cases \\ 
PPE & BANG.BANG.CON \\
essential & corona \\ 
Covid-19 & Pandemic \\
FANTASIA & pandemic \\ 
CRABS & infected \\
vol.7 & asymptomatic \\ 
lockdowns & virus \\
Lockdown & coronavirus \\ \hline
\end{tabular}
\caption{
    We display the top 20 English words closest to the centroids of $C_1$ and $C_2$.
    The pattern of words assigned to $C_1$ being more specific and concerned with social distancing in particular, while 
    words assigned to $C_2$ are focused on pandemics more generally, is apparent.
    Because we do not perform any explicit topic modeling, some words are included in this list that do not correspond to COVID-19-related topics. They are close to the cluster centroids just because they display very similar intertemporal collective attention dynamics.
    Though individual words may change rank, this list is qualitatively insensitive to regeneration of clusters as new data becomes available. 
}
\label{tab:top20-english}
\end{table}
\medskip

\noindent
Words assigned to $C_1$ reflect immediate measures taken to prevent the spread of 
COVID-19, such as ``flatten'', ``distancing'', ``télétravail'' (telework), ``hospitalier'' (hospital),
``encerrado'' (closed), and ``evitar'' (to avoid).

In contrast, words assigned to $C_2$ include more conceptual words that describe people, agencies, institutions,
and concepts surrounding epidemics more generally, such as ``pandemic'', ``CDC'', ``epidemiologist'',
``l'épidémie'' (the epidemic), ``virus'', ``contagiado'' (contagious).
Words assigned to $C_2$ describe pandemics in general, while words assigned to $C_1$ describe quarantines and lockdowns in particular, and words particular to this pandemic (e.g., ``Hydroxychloroquine'').
Though we have not conducted a formal linguistic analysis to conclude that there are significant semantic
differences between words assigned to each cluster, these preliminary findings provide evidence that such a 
semantic difference does exist. 

\subsection{Death attribution by language}
Using the methodology described in Sec.\ \ref{sec:case-num-lang}, we aggregate country-wide infection case numbers
and bin them approximately by language, thus enabling analysis of new cases stratified by language.
Because the log word rank time series are nonstationary and the new case number time series are not 
scale-independent, we move to a percent-change based analysis of these data.
We analyze percent-change time series for the mean log rank timeseries of the top 20 words closest to $E[C_2]$
and the new case time series for each language.
We estimate a latent volatility statistic for each percent-change time series.
This statistic captures the latent variance of the time series at each point in time without destroying 
information through computation of a rolling variance.
\begin{figure}
\centering
\includegraphics[width=\columnwidth]{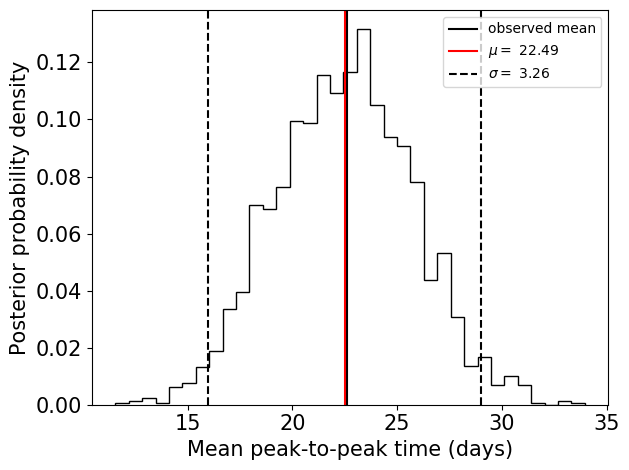}
\caption{
Mean distance between the peak latent volatility of percent-change in new cases and 
peak latent volatility of percent-change of closest $C_2$ word to $E[C_2]$ is approximately
$\mu = 22.49$ days, with a standard deviation of the mean given by 
$\sigma \approx 3.26$ days under the difference-of-Poissons model.
This model captures the middle third of the observed P2PD empirical cdf, but does not capture the tail behavior of this distribution.
}
\label{fig:p2p-mean}
\end{figure} 

Peaks in the latent volatility statistic indicate days on which the underlying time series 
exhibited large percent-changes in its value.
We measure the distance between (a) the peak latent volatility statistic of percent-change log word rank and (b) the
peak latent volatility statistic of percent-change new cases, termed the peak-to-peak distance (P2PD), for 
each language under study.
P2PD is a simple metric of the lag between fluctuations in social media attention to the initial 
Coronavirus outbreak and (usually positive) large fluctuations in new cases.
P2PD is greater than zero for all but four languages (Swedish, Urdu, Finnish, and Ukrainian) under study.

Observed values of P2PD are partially reproduced by a simple Poisson data generating process. 
We model the days at which peak volatility of each percent-change time series occurred
as being generated by a Poisson distribution with an unknown rate parameter.
P2PD is then modeled as the number of days between the peak day of new case volatility 
and the peak day of log word rank volatility.
Under this difference-of-Poissons model, P2PD is approximately 22.57 days with a estimated SEM of 3.26 days.
The observed cumulative distribution function (cdf) of P2PD data is 
not wholly contained within the posterior distribution of empirical cdfs generated by this model; 
the middle third of the distribution is reproduced by this model, but the tails are not.
\begin{figure}
\centering
	\includegraphics[width=\columnwidth]{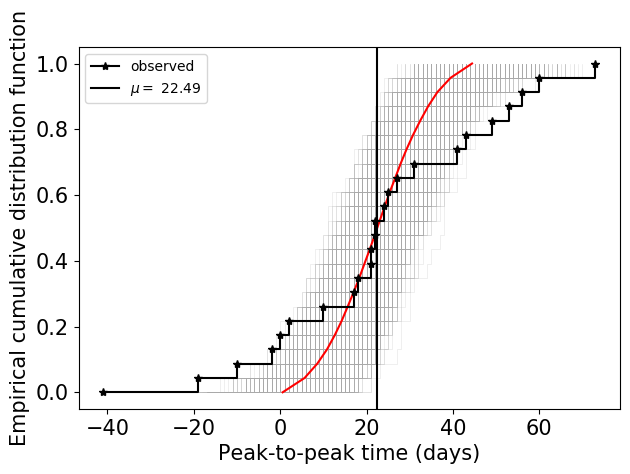}
	\caption{
		The observed empirical cdf of the P2PD data is reproduced by the posterior predictive 
		distribution of empirical cdfs of the Poisson model described in Sec.\ \ref{sec:case-num-lang}.
		The posterior mean $\mu$ of the P2PD data is displayed as a vertical black line.
		The middle third of the data is reproduced by the posterior distribution of empirical cdfs, but the tails of the distribution are not.
		We display the mean posterior empirical cdf in the red curve.
		}
		\label{fig:p2p-poisson}
\end{figure}
We display the distribution of estimated mean P2PD in Fig.\ \ref{fig:p2p-mean}
and empirical cdfs generated by the posterior predictive distribution of the difference-of-Poissons model in Fig.\ \ref{fig:p2p-poisson}.

\begin{figure}
    \centering
    \includegraphics[width=\columnwidth]{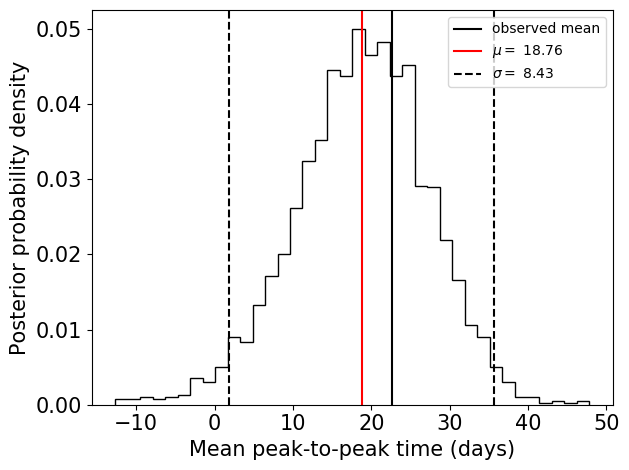}
    \caption{Distribution of posterior mean P2PD under the alternative, Dirichlet process Poisson mixture model.
    This model imposes regularization toward zero mean P2PD. Nonetheless, a $2\sigma$ uncertainty interval around the posterior mean P2PD still excludes zero.}
    \label{fig:p2p-mean-dp}
\end{figure}

\begin{figure}
    \centering
    \includegraphics[width=\columnwidth]{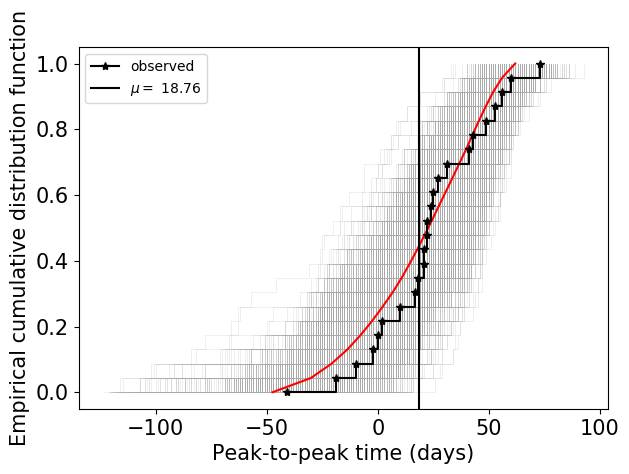}
    \caption{The empirical cdf of the P2PD data has high posterior probability under the Dirichlet process Poisson mixture model.
    We display the mean posterior empirical cdf in the red curve.
    }
    \label{fig:p2p-dp}
\end{figure}

A Dirichlet process Poisson mixture model, an alternative model that hypothesizes subpopulation heterogeneity in P2PD data, does accurately reproduce the observed empirical cdf of P2PD data, as we demonstrate in Figs.\ \ref{fig:p2p-mean-dp} and \ref{fig:p2p-dp}.
We discuss the specifics of this model in greater depth in Sec.\ \ref{sec:methods}.
The estimated distribution of posterior mean P2PD under this alternative model (18.76 $\pm$ 8.43 days) is lower and has higher variance than the estimated distribution of posterior mean (22.49 $\pm$ 3.26) under the difference-of-Poissons model.
Even though the alternative model imposes strong regularization toward zero mean P2PD, a two-standard deviation uncertainty interval for posterior mean P2PD does not contain zero. 

%% file: sections/conclusion.tex
Analyzing the behavior of words found in a random 10\% sample of all tweets between 
2019-09-01 and 2020-03-25, we find a distinct bilateral split in dynamics of words relating to the 
COVID-19 pandemic.
Though we have not performed a formal linguistic analysis, evidence suggests that 
words used to describe the initial reports of a Coronavirus outbreak in China differ semantically 
from words used later to describe the worldwide fight against the pandemic. This second cluster reflects discussion of specific measures, such as quarantine and
social distancing, currently being used to mitigate the spread of the virus and limit casualties.

The initial spike in collective attention to the Coronavirus in mid-January 2020, 
subsequently followed by a decay, is explained by the dynamics of the first cluster of 
words and not the second.
The mean number of days between peak volatility of percent change in first-cluster words and 
peak volatility of percent change in new case numbers (P2PD) is approximately 23 days, which 
is comparable to estimates of right-censored median time delay betweeen onset of COVID-19 and 
death \cite{linton2020incubation,world2020report} and median duration of viral shedding
\cite{zhou2020clinical}.
The observed distribution of P2PD is statistically reproduced by a simple difference-of-Poissons model 
when aggregating across all languages under study.
\medskip

This study is exploratory. We take care to not extrapolate from the current set of results without 
adequate caution.
First, we use only the top 1000 words in each language as ranked in April 2020 when compared with April 2019 \cite{alshaabi2020world}. 
This list of words is dynamic and may change our results either quantitatively or qualitatively. 
Second, all of our results are non-causal because we analyze the entirety of each time series
(word time series and new infection case numbers). 

Associations that we find should not be taken as causal, 
or even necessarily predictive, for two reasons. 
It is obvious that change in word usage rank on Twitter does not cause new cases of COVID-19.
Though it may be possible to use change in word usage rank to inform predictions of new case numbers,
we have not performed such forecasting ourselves and it is possible that these results will not hold in the 
future.
In addition, this time delay may be applicable only to COVID-19 and not necessarily other infectious diseases.
While we have attempted to control for nonstationarity and explicit time dependence by analyzing percent changes and their variance --- and not analyzing correlation between the nonstationary time series
of log word rank and new case numbers --- this does not mean that the association is not spurious and more extensive analysis of this association is warranted.

While it is suggestive that mean P2PD is comparable to estimates of time delay between COVID-19 onset and death,
we particularly hesitate to draw any conclusions from this observation, though it should be a target of
further theoretical and empirical study. 
We do not have subject-matter expertise in epidemiology and so will not offer speculation on this matter.
\medskip

There are several ways in which this study could be extended, for example by continuously updating words through time in order to test our methods' generalizability.
More importantly, the methodology can be applied to other infectious disease outbreak data to test our
hypothesis that changes in social media attention to epidemic-related words can provide a useful 
signal in predicting future new case volatility.
Future studies could also use more sophisticated clustering,
similarity search or latent volatility estimation methods
\cite{kastner2014ancillarity,dewhurst2020shocklet}.

%% file: sections/methods.tex
\subsection{Cluster number selection}\label{sec:cluster-number-selection}
We clustered the log word rank time series $\log_{10} \frac{1}{r_t}$ using the minibatch $k$-means clustering (KMC) algorithm \cite{sculley2010web}.
Before clustering, we normalized the time series so that clusters would not form purely based on the average rank of 
each word.
The functional form of the normalization was $\log_{10} \frac{1}{r_t} 
\mapsto
\frac{\log_{10} \frac{1}{r_t} - \mu}{\sigma}$, where
$\mu = \frac{1}{T}\sum_{t=1}^T \log_{10}\frac{1}{r_t} $ and 
$\sigma^2 = \frac{1}{T}\sum_{t=1}^T (\log_{10} \frac{1}{r_t}  - \mu)^2$.

\begin{figure}
\centering
	\includegraphics[width=\columnwidth]{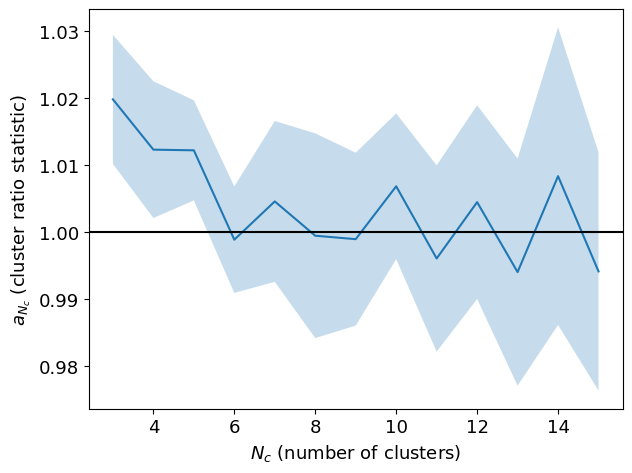}
	\caption{
		Mean and bootstrapped two-standard deviation uncertainty intervals for ratio statistic used in choosing number of clusters
		in minibatch $k$-means algorithm.
		}
	\label{fig:elbow}
\end{figure}

\begin{figure*}
\centering
	\includegraphics[width=\textwidth]{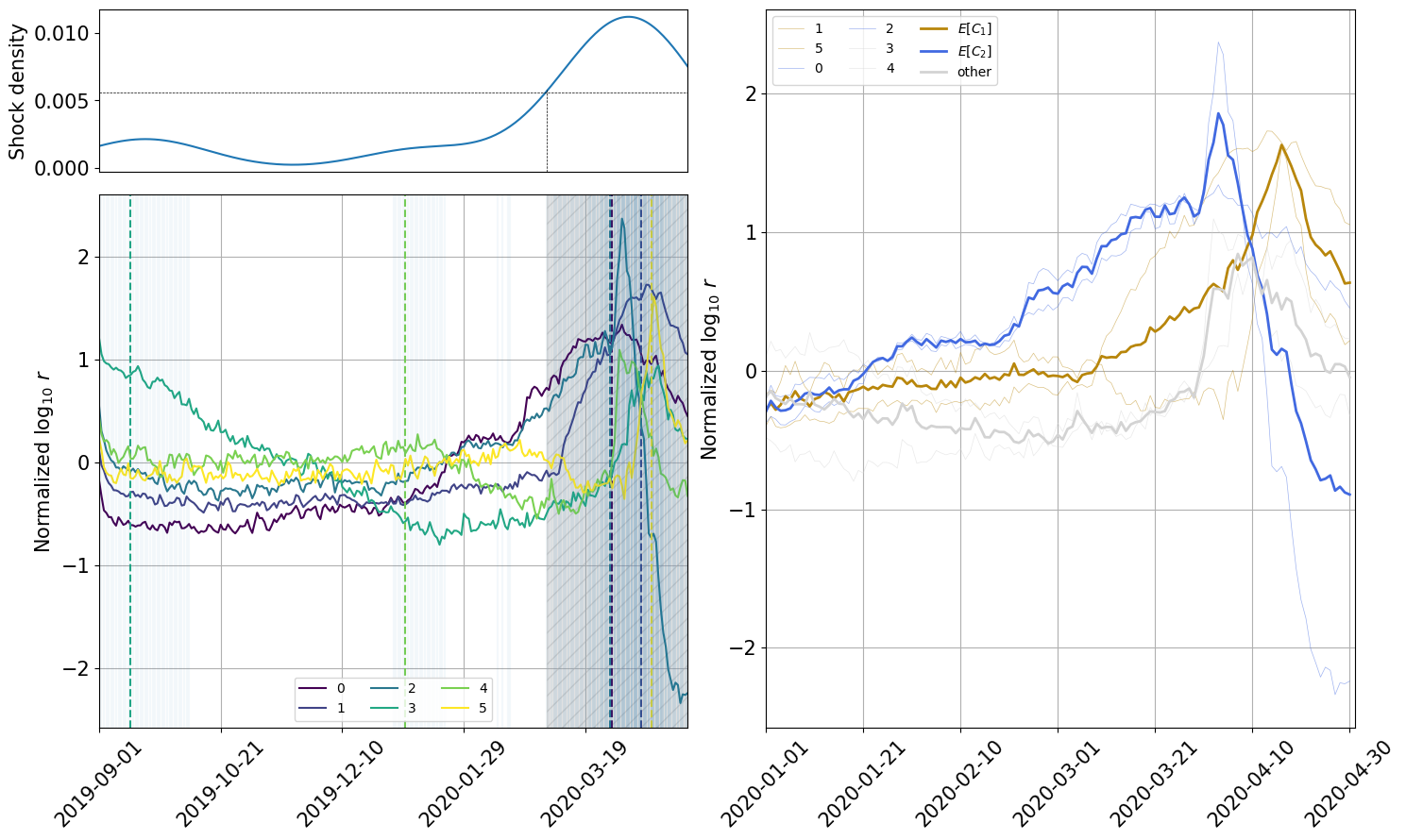}
	\caption{
		Using the algorithm detailed in Sec.\ \ref{sec:cluster-number-selection}, 
		we find $k^* = 6$ clusters of log rank word time series during the time period
		2019-09-01 to 2020-03-25.
		We do not label the clusters with informative labels because clusters are unique only up to a permutation of labels.
		}
		\label{fig:all-ts-clusters}
\end{figure*}

We chose the number of clusters $k^*$ using the following algorithm \footnote{
This algorithm is a modified version of an unpublished algorithm 
\cite{daveisossum_2017}.
}.
For each of $N$ independent trials, we fit a minibatch KMC model for each of $k = 1,...,15$ clusters.
For each of these clusters in each independent trial, we recorded the average Euclidean distance 
of the set of all time series from the closest cluster centroid. 
We denote this error metric by $\ell_{n,k}$. 
We then computed a ratio-of-ratios statistic, 
$a_{n,k} = \frac{\ell_{n,k+1}}{\ell_{n,k}} \Big / \frac{\ell_{n,k}}{\ell_{n,k-1}}$,
and selected the number of clusters as 
$k^* = \argmin \left\{k - 1:\ a_k \leq 1 \right\}$,
where we have put $a_k = \frac{1}{N}\sum_{n=1}^N a_{n,k}$.
We display bootstrapped single standard deviation confidence intervals around $a_k$ in
Fig.\ \ref{fig:elbow}.
This algorithm returned the number of clusters $k^* = 6$.
We then collapsed the number of clusters based on dynamic behavior. 
Panel (b) of Fig.\ \ref{fig:all-ts-clusters} displays each cluster centroid. 

We extracted time windows where cluster centroids displayed increasing rates of increase followed by decreasing rates of decrease using the discrete shocklet transform \cite{dewhurst2020shocklet}. 
This dynamic behavior corresponds with increased collective attention to words and topics associated with that cluster centroid followed by decreased collective attention. 
Each time window is composed of one or more time points. To aggregate this ``cusplike'' behavior, we placed a Gaussian kernel around each extracted time point and analyzed the resulting function, which we display in panel (a) of Fig.\ \ref{fig:all-ts-clusters} and which we denote by $S(t)$.
We considered cluster centroids to be temporally-relevant to our analysis of COVID-19 language dynamics if their maxima occurred in time intervals where $S(t)$ was equal to at least half of its maximum.
This condition was satisfied for four of the clusters during one time window,
2020-03-03 to 2020-04-30.
The maxima of these four clusters neatly partition into two groups.
One group has maxima that occur in late March and the other has maxima that occur in mid April. We combine the four clusters into two aggregated clusters based on this criterion and label the aggregate clusters $C_2$ and $C_1$ respectively. 
We display $C_1$ and $C_2$ in panel (c) of Fig.\ \ref{fig:all-ts-clusters}.

We used tweets authored both before and during the COVID-19 pandemic to generate the clusters, 
so the centroids are relatively flat before the 
initial coronavirus reports (late December 2019) and some exhibit
periodic behavior. 
The magnitude of the horizontal axis is lower than in 
Figs.\ \ref{fig:cluster-decomp-english},
\ref{fig:cluster-set0}, and \ref{fig:cluster-set1} because 
here we display only the cluster centroids, which necessarily have moderated fluctuations compared to the more extreme 
cluster elements displayed in other figures.

\subsection{Case number attribution by language}\label{sec:case-num-lang}
To associate country-level case number changes with languages, we performed a one-to-one lossy mapping of 
country to dominant language spoken in that country.
Using data from the Australian federal government's Department of 
Social Services, we truncated the list of languages spoken in each country to the most prevalent language 
in each country.
While this mapping is crude and eliminates subtleties of intranational language diversity (e.g.,
Switzerland is mapped solely to German, while French, Italian, and Romansh are dropped), it allowed us to 
reverse the direction of this mapping and assign to each language the number of new cases equal to the sum of 
new cases in each country for which the language is the primary language.
We obtained new case numbers from the European Center for Disease Control and Prevention.
\medskip

\subsection{Volatility characterization}\label{sec:volatility-characterization}
We move to a percent-change approach in our joint analysis of new case numbers and log rank word time series
because log rank word time series are nonstationary (they are only wide-sense stationary in our analysis because we
normalize them to have intertemporal zero mean and unit variance) and new case number time series 
are not scale-independent.
We define the percent-change time series as $y_t \equiv \log \frac{x_t}{x_{t-1}}$, where $x_t \in \{ \text{log rank word time series}, \text{new case time series}\}$.
Instead of analyzing $y_t$, an unbounded random variable, we instead analyze the variance of 
$y_t$, denoted $s_t$,
by estimating a standard Bayesian stochastic volatility model \cite{jacquier2002bayesian,gelman2013bayesian}.
We hypothesize that the latent log-variance $s_t$ evolves according to 
\begin{equation}
	s_t \sim \text{Normal}(s_{t-1}, v^2),\ s_0 \sim \text{Normal}(0, 1).
\end{equation}
We place a weakly informative prior on the standard deviation of the increments of this process, 
$v \sim \text{LogNormal}(0, 1)$.
The percent change is then modeled as 
\begin{equation}
	y_t \sim \text{Normal}(\mu, \exp (s_t/2)),
\end{equation}
where we include a tight zero-centered prior for the mean percent change,
$\mu \sim \text{Normal}(0, 0.01)$.
We fit this model using stochastic variational inference with a diagonal normal guide
(variational posterior) \cite{hoffman2013stochastic}.
We conduct optimization using the Adam optimizer with a learning rate of 0.05 and run the optimizer for a total of 1500 iterations 
\cite{kingma2014adam}.
We conduct our subsequent volatility analysis using draws from the optimized guide distribution.
\medskip

In addition to estimating the mean P2PD, we model P2PD with a null, simple model and an alternative, more complex model.

In the null model, 
we suppose that the number of days to peak in average latent volatility in each of the percent change time series
is given by $N_{s} \sim \text{Poisson}(\lambda)$, where we place a weakly informative prior on the rate parameter,
$\lambda \sim \text{LogNormal}(0, 1)$. 
We sample from this Poisson model for each of the percent change time series, 
and then model P2PD as the difference in these Poisson rvs (known as a Skellam distribution).
We use the No U-Turn Sampler (NUTS) algorithm to sample from the posterior \cite{hoffman2014no}, sampling from one chain for 500 iterations of warmup followed by 2500 iterations of sampling.
We display draws from the posterior predictive distribution of empirical cdfs in Fig.\ \ref{fig:p2p-poisson},
along with the observed empirical cdf of the P2PD data.
This null model does not adequately capture the shape of the empirical cumulative distribution function (cdf) of the observed P2PD data.
Though it does capture the distribution of the middle third of the observations well,
the tails of the observed P2PD are heavier than predicted by this model.
In particular, the tails of the observed empirical cdf lie outside of the distribution of posterior empirical cdfs generated by this model.

We then move to a Dirichlet process Poisson mixture, a more expressive alternative model.
This model hypothesizes that the data are generated by a possibly-infinite mixture of Poisson probability distributions. (More technical definitions can be found in a variety of references 
\cite{sethuraman1994constructive,neal2000markov,teh2005sharing,dahl2006model}.)
The mixture weights are drawn as 
$w_i = \beta_i \prod_{j < i} \beta_j$ for each $i = 1,...,N$.
Each $\beta_i$ is drawn independently as
$\beta_i \sim \text{Beta}(1, 1) \equiv \text{Uniform}(0, 1)$.
Each component Poisson distribution has rate parameter distributed
independently as $\lambda_i \sim \text{LogNormal}(0, 1)$.
We model each component of P2PD data (peaks in new case volatility and peaks in $C_2$ volatility) using this model, and then again model P2PD as the difference in these random variables.
In practice the Dirichlet process must be truncated to a finite number of components $N$.
We truncate to $N = 3$ as there is not a substantial difference in the distributions of empirical cdfs for $N = 6, 9, 12$. (We present more details in Appendix \ref{app:dp-poisson}.)
We again fit this model using NUTS, this time sampling from one chain for 1000 iterations of warmup followed by 2500 iterations of sampling.

This model describes the entire empirical distribution of observed data 
well, as the observed empirical cdf lies entirely within the distribution of posterior empirical cdfs generated by this model.

We chose the Dirichlet process Poisson mixture model over a more conventional model of overdispersed count data, such as a negative binomial model, because we do not believe that the mechanistic interpretation of a negative binomial model (number of failures observed before a given fixed number of successes) applies in the context of counting number of days from a reference date until peak volatility.
It is unclear what a ``success'' or ``failure'' would be in this context. On the other hand, the Poisson mixture model has a clear mechanistic interpretation: there is subpopulation heterogeneity among languages and countries grouped by language, but within each subpopulation the number of days from reference date until peak volatility occurs with a constant subpopulation-specific mean rate.

%% file: sections/acknowledgements.tex
The authors are grateful for the computing resources provided by the Vermont Advanced Computing Core 
and financial support from the Massachusetts Mutual Life Insurance Company and Google.

%% file: sections/supplementary.tex
\onecolumngrid
\section{DP Poisson mixture results}\label{app:dp-poisson}
\begin{figure}[!htp]
    \centering
    \includegraphics[width=\columnwidth]{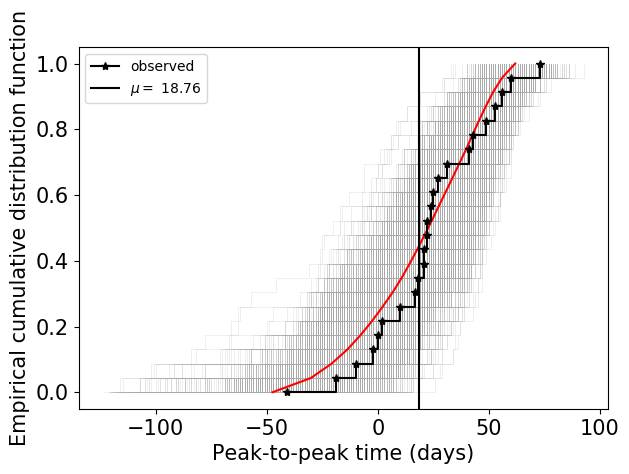}
    \caption{Posterior distribution of empirical cdfs of the Dirichlet process Poisson model with number of components truncated to $N = 3$.}
    \label{fig:dp-n3}
\end{figure}

\begin{figure}[!htp]
    \centering
    \includegraphics[width=\columnwidth]{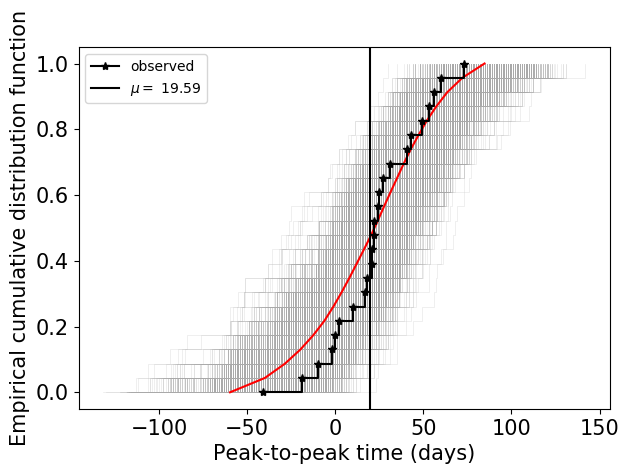}
    \caption{Posterior distribution of empirical cdfs of the Dirichlet process Poisson model with number of components truncated to $N = 6$.}
    \label{fig:dp-n6}
\end{figure}

\begin{figure}[!htp]
    \centering
    \includegraphics[width=\columnwidth]{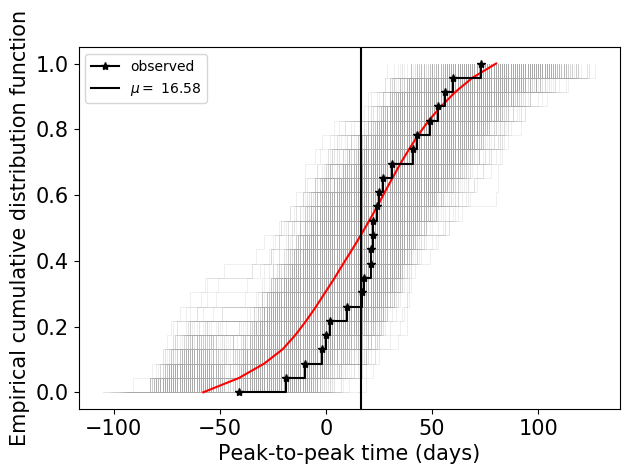}
    \caption{Posterior distribution of empirical cdfs of the Dirichlet process Poisson model with number of components truncated to $N = 9$.
    The region of increased density between -50 and -100 peak-to-peak time is a sampling artifact.}
    \label{fig:dp-n9}
\end{figure}

\begin{figure}[!htp]
    \centering
    \includegraphics[width=\columnwidth]{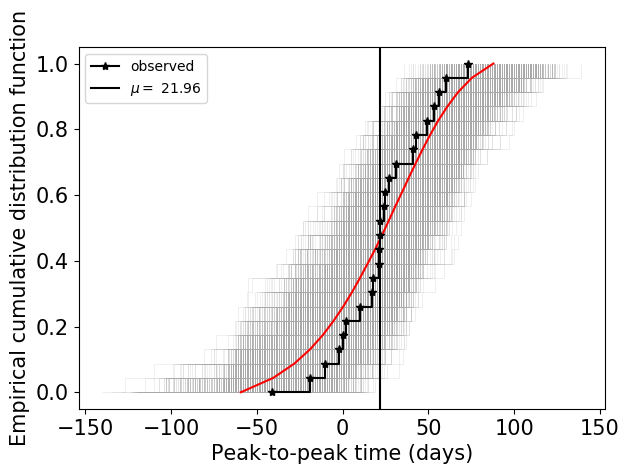}
    \caption{Posterior distribution of empirical cdfs of the Dirichlet process Poisson model with number of components truncated to $N = 12$.}
    \label{fig:dp-n12}
\end{figure}

The Dirichlet process Poisson mixture model is able to capture heterogeneity in the distribution of P2PD.
We display posterior distributions of empirical cdfs of Dirichlet process Poisson mixtures. We sampled from four different realizations of this model with number of components truncated to $N = 3, 6, 9, 12$.
We sampled from each model using the NUTS sampler, 1000 iterations of burnin, and 2500 iterations of sampling. 
Changing $N$ did not substantially alter the fit of the models, as we display in Figs.\ \ref{fig:dp-n3} - \ref{fig:dp-n12}.
Because of this, we choose the most parsimonious of these models and set $N = 3$ for analysis.

